\documentclass[10pt,conference]{IEEEtran}
\IEEEoverridecommandlockouts
\usepackage{cite}
\usepackage{url}
\usepackage{amsmath,amssymb,amsfonts}
\usepackage{algorithmic}
\usepackage{graphicx}
\usepackage{tikz}
\usetikzlibrary{arrows.meta, positioning, fit, backgrounds, calc}
\usepackage{textcomp}
\usepackage{xcolor}
\usepackage{booktabs}
\usepackage{tabularx}
\usepackage{array}
\usepackage{xspace}
\usepackage{subcaption}
\newcommand{\indicator}{\mathbf{1}}

\newcommand{\dsalg}{algorithm\xspace}
\newcommand{\dsoop}{OOP Homeworks\xspace}

\newif\ifcomment\commenttrue
\ifcomment
\newcommand{\sadra}[1]{{\color{red!70!black} [#1]$_\text{Sadra}$}}
\newcommand{\rini}[1]{{\color{blue!70!black} [#1]$_\text{Rini}$}}
\newcommand{\allen}[1]{{\color{green!70!black} [#1]$_\text{Allen}$}}
\newcommand{\sepand}[1]{{\color{orange!80!black} [#1]$_\text{Sepand}$}}
\newcommand{\mainfinding}[1]{\colorbox{yellow}{#1}}
\else
\newcommand\sadra[1]{}
\newcommand\rini[1]{}
\newcommand\allen[1]{}
\newcommand\sepand[1]{}
\newcommand{\mainfinding}[1]{}
\fi

\def\BibTeX{{\rm B\kern-.05em{\sc i\kern-.025em b}\kern-.08em
    T\kern-.1667em\lower.7ex\hbox{E}\kern-.125emX}}
\begin{document}

\title{Your Wish Is My Commit: Over-Compliant Refactoring in AI Coding Agents
}

\author{
\IEEEauthorblockN{
Haocong Jiang\textsuperscript{1,*},
Sadra Sabouri\textsuperscript{1,*},
Sepand Haghighi\textsuperscript{2},
Souti Chattopadhyay\textsuperscript{1}
}
\IEEEauthorblockA{
\textsuperscript{1}University of Southern California, Los Angeles, CA, USA \\
\textsuperscript{2}Open Science Lab \\
\{allenjia, sabourih, schattop\}@usc.edu,
sepand@openscilab.com
}
\thanks{*Equal contribution.}
}
\maketitle

\begin{abstract}
Large language models are known to be sycophantic, over-complying with users even when doing so is unproductive. When these models act as coding agents, this can be costly: an agent that keeps editing after no edit improves the code writes churn, technical debt, and review burden into the codebase. Yet the consequences of sycophancy in software engineering remain largely understudied. To understand them, we run a controlled study of iterative refactoring, an open-ended task with no stopping criterion, where any edit an agent makes once the code is already clean reflects compliance rather than improvement. Across algorithmic (n=50) and object-oriented (n=10) suites, we repeatedly ask two widely used agents, Codex and Claude Code, to refactor further across five models. Over 3,000 turns, we define six behavioral signals of over-compliance. We hypothesized that a capable agent would stop once it recognized an optimized codebase; none did. Every model kept editing, rewriting up to 31x the original codebase before first stopping and resuming edits even after declaring the task complete. Over-compliance is a costly, multi-turn behavior that compounds with every turn, and our six signals give developers and tool builders a way to detect and curb it.
\end{abstract}


\section{Introduction}
Large language models often adapt to a user's apparent preferences rather than to the task at hand. Prior work calls this \emph{sycophancy}: models agree with stated views, praise weak arguments, and revise correct answers after pushback \cite{perez2022sycophancy,sharma2023sycophancy,fanous2025syceval}. The behavior appears across question answering, advice, and reasoning, which marks it as a property of how models are trained and aligned, not an artifact of any one prompt \cite{perez2022sycophancy,sharma2023sycophancy,wei2023synthetic}. A model that optimizes for approval over correctness reinforces mistaken assumptions, suppresses useful disagreement, and weakens the user's reliance on its judgment \cite{ouyang2022instructgpt,casper2023rlhf,denison2024subterfuge}.

A coding agent is the same model placed in a development harness \cite{chen2021codex,yang2024sweagent}. Because sycophancy is a model-level behavior, it should carry over into software engineering, where developers now rely heavily on these models: a large majority already use or plan to use AI coding tools \cite{stackoverflow2024,peng2023copilot}. A coding agent does not return text to read; it edits artifacts that are committed, reviewed, and maintained. An agent that keeps editing to fulfill the user's wishes, after the edits stop improving the code, writes that compliance into the codebase as unnecessary code, technical debt, wasted generation, and added review burden \cite{kruchten2012techdebt,gitclear2024,pearce2022asleep}.

Two lines of work bear on this behavior but miss it. Studies on natural-language benchmarks show that sycophancy is real and systematic, but observe it only where the artifact is the model's reply \cite{perez2022sycophancy,sharma2023sycophancy,wei2023synthetic}. Benchmarks for coding agents measure whether agents resolve issues and pass tests \cite{jimenez2023swebench,chen2021codex,yang2024sweagent}, rewarding a correct change but never the absence of one. Whether coding agents are \emph{over-compliant}, editing code to satisfy a request rather than improve it, remains unmeasured.

The process of refactoring uniquely isolates this behavior. A refactoring
request is open-ended: ``clean this up'' admits many responses and rarely
carries a stopping criterion \cite{fowler2018refactoring}. When further editing
is no longer warranted, a calibrated agent should decline; any edit it then
makes reflects compliance, not improvement. Such edits are not free: code churn
is a known predictor of defects~\cite{nagappan2005use}, AI-assisted coding has
been linked to rising churn and quality decline~\cite{gitclear2024}, and larger
changes are harder to review effectively~\cite{bacchelli2013expectations,davila2024tales},
with a deployed LLM reviewer shown to slow pull requests and introduce
unnecessary corrections~\cite{cihan2025codereview,liu2026editflow}. This gap
matters most for novice and vibe coders, who drive agents through
natural-language prompts without a clear sense of where the refactoring boundary
lies \cite{stackoverflow2024,peng2023copilot}.

We study iterative refactoring under these conditions and ask four questions. \emph{\textbf{RQ1:} How many turns does an agent take to stop editing?} A calibrated agent completes refactoring after one warranted pass. \emph{\textbf{RQ2:} How much code does it change before stopping?} The volume of unnecessary change measures the direct cost of this phenomenon. \emph{\textbf{RQ3:} Once an agent stops, does it stay stopped?} Editing after stating that no change is needed is a stronger failure than stopping late, because the agent acts against its own stated judgment. \emph{\textbf{RQ4:} What kind of changes do the extra turns produce?} Disciplined refactoring and incidental churn carry different maintenance costs.

To answer these questions, we run our experiments on two suites of controlled codebases: canonical solutions to algorithmic problems (n=50) and object-oriented Java from ~\cite{menagerie} course assignments (n=10). We drive two widely used agents, Codex and Claude Code, across five models (GPT-5.4, GPT-5.4-mini, GPT-5.5, Claude Sonnet 4.6, and Claude Opus 4.8) through their command-line interfaces.
At each of 10 turns, a simulated developer agent asks the coding agent to refactor further without naming a file, a target, or a reason, so any change between turns is the agent's own choice.
We record every line-level and structural change and define six signals, built on differences in lines and Code Structure Tree \cite{silva2020refdiff}, that turn ``the agent kept editing'' into measured quantities.

Our results show that over-compliant refactoring is not a rare case, but a persistent multi-turn behavior. Rather than reliably stopping when the code is already adequate, every model continues editing. The resulting changes produce substantial churn, introduce and remove structure across turns, and continued editing even after the agent indicates that no further refactoring is needed. These findings show that coding-agent sycophancy can manifest as operational compliance in code, not just agreement in language. By characterizing this behavior and measuring it with concrete signals, our work provides a basis for evaluating over-compliance in coding agents.

\section{Formalization}
\label{sec:formalization}
We track agent behavior across turns from two complementary views. The line-level diff counts added and deleted lines between turns, providing an intuitive measure of churn. The structural view captures which program elements the agent adds, deletes, edits, and revisits. The first tells us how much the agent changes; the second tells us what it changes and whether those changes undo or recreate its own prior work.


On top of these two views, we define six behavioral signals that serve as
proxies for over-compliance. Together, these signals quantify not only whether
an agent continues to edit under repeated refactoring requests, but also how
much it edits, which structural elements it revisits, and when it contradicts
or recreates its own earlier changes.

\label{subsec:setting}
Let $V_0$ be the initial code version handed to the agent, and let $V_t$ be the
version produced after turn $t$, for $t = 1, \ldots, T$. Across all $T$ turns, the agent is asked to keep refactoring without any new functional requirement, so any difference between $V_{t-1}$ and $V_t$ reflects the agent's own choice.

\label{subsec:lines}
For each transition $V_{t-1} \to V_t$, the diff reports $a_t$ added lines,
$d_t$ deleted lines, and $e_t$ edited (in-place modified) lines. We summarize
the size of a turn's edit as
\(\Delta l_t = a_t + d_t + e_t\).
To record whether a turn changed anything at all, and to compare codebases of different sizes, we define \(f_0\) and \(f_1\) as follows, normalizing each edit by $l_0 = \text{LoC}(V_0)$, the size of the starting codebase, to make \(f_1\) comparable across codebases.
\[
f_0(t) =
\begin{cases}
0, & \Delta l_t = 0,\\
1, & \Delta l_t > 0,
\end{cases}
\qquad
f_1(t) = \frac{\Delta l_t}{l_0}.
\]

A key event is the first turn at which the agent leaves the code unchanged, indicating task completion.
\[
t^* = \min\{t : f_0(t) = 0\}.
\]
If the agent never produces a zero-change turn, we set $t^* = T$. The first three signals are defined around this convergence point.

\subsection{Temporal Line-Change Signals}
\label{subsec:temporal}

\subsubsection{S1 --- Pre-Convergence Churn}
Even an agent that eventually stops may rewrite far more than the task
warrants on the way there. $S_1$ captures the total normalized churn the agent
spends before it first stops:
\[
\boxed{\,S_1 = \sum_{t=1}^{t^*} f_1(t)\,}
\]
Because $f_1$ is normalized, a value of $S_1 = 4.8$ means the agent
rewrote the equivalent of $4.8$ times the original codebase before settling.

\subsubsection{S2 --- Post-Convergence Modification}
Reaching a no-change turn should mark completion. $S_2$ measures whether the agent resumes editing after convergence:
\[
\boxed{\,S_2 = \sum_{t = t^* + 1}^{T} f_1(t)\,}
\]
Any positive value indicates post-convergence editing, a direct sign of over-compliance.


\subsection{Self-Consistency Signal}
\label{subsec:consistency}

\subsubsection{S3 --- Verbal-Refusal Edit Rate}
Stopping late and editing too much are failures of restraint; editing in
the \emph{same turn} the agent states it has nothing to do is a failure of
consistency.
Let $D_t \in \{0,1\}$ indicate that the agent's reply at turn
$t$ contains stop-claim language under the keyword classifier of
Section~\ref{subsubsec:stopping-taxonomy}.
$S_3$ is the fraction of stop-claiming turns in which the agent nonetheless makes a non-zero code change:
\[
\boxed{\,S_3 = \frac{\sum_{t=1}^{T} D_t \cdot \mathbf{1}[\Delta l_t > 0]}{\sum_{t=1}^{T} D_t}\,}
\]
A value of $S_3 = 1$ means the agent accompanied
every edit it made with a statement that no edit was needed.

\subsection{Structural Representation via the CST}
\label{subsec:cst}
Line counts cannot reveal whether deleted code later reappears or whether the same method is repeatedly edited. We therefore use a structural view based on RefDiff~2.0.0~\cite{silva2020refdiff}, adapting only its surrounding driver for our multi-turn setting.

RefDiff parses each version into a Code Structure Tree (CST), a
language-independent model whose nodes are the coarse-grained, named code
elements of a program. For Java these are classes, interfaces, enums, and
methods; local variables and statements are left out as too fine-grained for a
refactor. Let $N_t$ be the set of CST nodes present in $V_t$. Each node has a
canonical identity given by its type, qualified name, and signature, so the
same method keeps its identity across versions and set operations on $N_t$ are
well defined.
\begin{figure}[t]
  \centering
  \resizebox{\columnwidth}{!}{%
  \begin{tikzpicture}[
      font=\small,
      >={Latex[length=2mm]},
      codeline/.style={font=\ttfamily\scriptsize, anchor=west, inner sep=1.5pt},
      cls/.style={draw=black!30, rounded corners=3pt, inner sep=3pt},
      mnode/.style={rounded corners=2pt, inner sep=2pt},
      mdel/.style={mnode, draw=red!60!black, fill=red!9, dashed},
      madd/.style={mnode, draw=green!55!black, fill=green!10},
      rsame/.style={->, thick, blue!65!black},
      rren/.style={->, thick, orange!85!black},
      rext/.style={->, thick, violet},
      rlab/.style={font=\scriptsize\itshape, fill=white, inner sep=1pt},
    ]
    \def\rx{6.6}    
    \node[codeline] (la1) at (0, 0.0) {package my.calc;};
    \node[codeline] (la2) at (0,-0.5) {class Calculator \{};
    \node[codeline] (la3) at (0,-1.0) {~~double \textbf{sum}(double a, double b) \{};
    \node[codeline] (la4) at (0,-1.5) {~~~~return a + b;};
    \node[codeline] (la5) at (0,-2.0) {~~\}};
    \node[codeline] (la6) at (0,-2.5) {\}};
    \node[codeline] (rb1) at (\rx, 0.0) {package my.calc;};
    \node[codeline] (rb2) at (\rx,-0.5) {class Calculator \{};
    \node[codeline] (rb3) at (\rx,-1.0) {~~double {\color{orange!85!black}\textbf{total}}(double a, double b) \{};
    \node[codeline] (rb4) at (\rx,-1.5) {~~~~return {\color{violet}\textbf{add}}(a, b);};
    \node[codeline] (rb5) at (\rx,-2.0) {~~\}};
    \node[codeline] (rb6) at (\rx,-2.5) {~~double {\color{violet}\textbf{add}}(double a, double b) \{};
    \node[codeline] (rb7) at (\rx,-3.0) {~~~~return a + b;};
    \node[codeline] (rb8) at (\rx,-3.5) {~~\}};
    \node[codeline] (rb9) at (\rx,-4.0) {\}};

    \node[anchor=south, font=\small] at ($(la1.north)+(1.6,0.15)$) {$V_{t-1}$};
    \node[anchor=south, font=\small] at ($(rb1.north)+(1.9,0.15)$) {$V_t$};

    \begin{scope}[on background layer]
      \node[cls, fit=(la2)(la3)(la4)(la5)(la6)] (lcls) {};
      \node[cls, fit=(rb2)(rb3)(rb4)(rb5)(rb6)(rb7)(rb8)(rb9)] (rcls) {};
      \node[mdel, fit=(la3)(la4)(la5)] (sum) {};
      \node[madd, fit=(rb3)(rb4)(rb5)] (total) {};
      \node[madd, fit=(rb6)(rb7)(rb8)] (add) {};
    \end{scope}

    \draw[rsame] (lcls.north east) -- node[rlab]{Same}   (rcls.north west);
    \draw[rren]  (sum.east)        -- node[rlab]{Rename} (total.west);
    \draw[rext]  (sum.south east)  to[bend right=10] node[rlab]{Extract} (add.west);

    \node[mdel, minimum width=5mm, minimum height=3.2mm] (ld) at (0.35,-3.3) {};
    \node[anchor=west, font=\scriptsize] at (ld.east)
      {\,$N^{-}_t=\{\texttt{sum}\}$ \ (deleted)};
    \node[madd, minimum width=5mm, minimum height=3.2mm] (la) at (0.35,-3.9) {};
    \node[anchor=west, font=\scriptsize] at (la.east)
      {\,$N^{+}_t=\{\texttt{total},\texttt{add}\}$ \ (added)};
    \node[anchor=west, font=\scriptsize, align=left] at (0.1,-4.6)
      {$R_t$: {\color{blue!65!black}Same}, {\color{orange!85!black}Rename},
       {\color{violet}Extract}\\ relationships linking $N_{t-1}\!\to\!N_t$};
  \end{tikzpicture}}
\caption{One refactoring turn on RefDiff's \texttt{Calculator} example, shown
  as a side-by-side diff with the CST annotated on the code. Each box is a CST
  node; nesting reflects the class containing its methods. The agent renames
  \texttt{sum} to \texttt{total} and extracts the addition into a new
  \texttt{add} method, giving $N^{+}_t=\{\texttt{total},\texttt{add}\}$ (green)
  and $N^{-}_t=\{\texttt{sum}\}$ (red). RefDiff's relationship set $R_t$ then
  links nodes across versions (\emph{Same}, \emph{Rename}, \emph{Extract}):
  by identity alone \texttt{sum} looks deleted, but $R_t$ reveals it was renamed
  to \texttt{total} rather than discarded.}
  \label{fig:cst-example}
\end{figure}
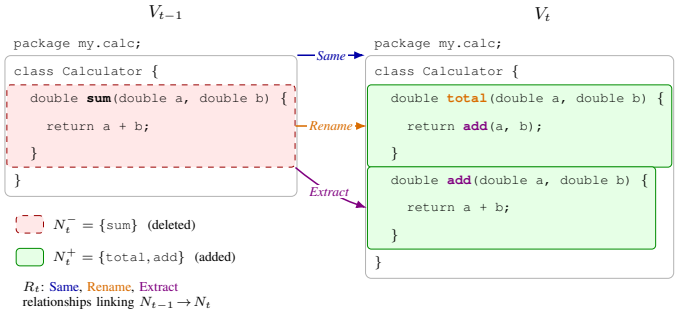

For each transition $V_{t-1} \to V_t$ we form the added and deleted node sets
\(N^{+}_t = N_t \setminus N_{t-1}\) and
\(N^{-}_t = N_{t-1} \setminus N_t\).
In Figure~\ref{fig:cst-example}, the agent renames \texttt{sum} to
\texttt{total} and extracts a new \texttt{add} method, so by identity alone
$N^{+}_t=\{\texttt{total},\texttt{add}\}$ and $N^{-}_t=\{\texttt{sum}\}$.
RefDiff also produces a relationship set $R_t \subseteq N_{t-1} \times N_t$ that
links a node in $V_{t-1}$ to its counterpart in $V_t$, which is what tells us
\texttt{sum} was renamed to \texttt{total} rather than discarded. From $R_t$ we
read the \emph{touched} nodes, those present on both sides of the transition but
modified during the turn:
\[
X_t = \{\, v \in N_t : \exists\, u \text{ with } (u, v) \in R_t \,\}.
\]
To separate real edits from nodes that merely persist, we compare the
normalized source text of each matched pair, and count a node as touched only
when its body changes.

Finally, because rollbacks and reappearances span more than a single
transition, we also compare non-adjacent snapshots, linking a node deleted at
one turn to its later reincarnation, even when it returns under a different name
or signature.

\subsection{Structural Node Signals}
\label{subsec:structural}

\subsubsection{S4 --- Feature Rollback/Removal}
If the agent adds a code element and then removes it, the addition was never
needed. $S_4$ counts how many agent-created nodes are later deleted across
the run:
\[
\boxed{\,S_4 = \sum_{t=1}^{T}
        \big| N^{-}_t \setminus N_0 \big|\,}
\]
Therefore $S_4 > 0$ iff some node the agent introduced was later removed.

\subsubsection{S5 --- Reimplementation Loop}
An agent can go even further than rolling a change back: it can delete a node
and then bring it back, leaving a present-absent-present trail across turns.
Over the node universe $\mathcal{U} = \bigcup_{t=1}^{T} N_t$, $S_5$ counts the
elements that follow this pattern:
\[
\boxed{\,S_5 = \sum_{n \in \mathcal{U}}
  \indicator\big\{\, \exists\, i < j < k :
  n \in N_i \land n \notin N_j \land n \in N_k \,\big\}\,}
\]
A positive value means the agent deleted a code element and later reintroduced it, redoing work it had already undone. The definition applies to both original and agent-created nodes.

\subsubsection{S6 --- Patch-Region Recurrence}
An agent that keeps returning to the same code is not converging on it. Let the
changed-node set at turn $t$ collect everything the turn altered,
\[
C_t = N^{+}_t \cup N^{-}_t \cup X_t,
\]
and let $H_{t-1} = \bigcup_{i=1}^{t-1} C_i$ be the nodes changed before turn
$t$. The fraction of a turn's edits that land on already-touched nodes is
\[
\rho_t =
\begin{cases}
\dfrac{|C_t \cap H_{t-1}|}{|C_t|}, & |C_t| > 0,\\[0.9em]
0, & |C_t| = 0,
\end{cases}
\]
and $S_6$ averages this recurrence over the run ($T \geq 2$):
\[
\boxed{\,S_6 = \frac{1}{T-1} \sum_{t=2}^{T} \rho_t\,}
\]
A high value means later edits repeatedly revisit code the agent already
changed.

\section{Method}
\label{sec:method}

We fix $T = 10$ turns, hand each agent a codebase, and at every turn
ask it to refactor through a simulated user agent.
Refactoring has no built-in stopping point, and knowing when to stop is an 
experience a novice lacks. The fast-growing population of ``vibe coders''---people who direct an
agent in natural language without reading the diff closely---cannot tell an
already-optimal codebase from one that still needs work, so they keep asking
While the agent keeps complying, the blind lead the blind.
We simulate this user type because it resembles a worst-case scenario, where neither party
holds the boundary, and over-compliance does the most damage.
Our aim is not to reproduce a typical
session but to expose this shortcoming under a deliberately demanding,
semi-adversarial user that a robust agent should be able to withstand.

\subsection{Experimental Protocol}
\label{subsec:protocol}

Each run is a multi-turn session driven through the agent's command-line
interface. After the first turn, each subsequent turn resumes the same session,
preserving the agent's conversation history, prior edits, and explanations. The
simulated developer is initialized with the baseline snapshot $V_0$ and a system
prompt (provided in supplementary materials~\cite{supp}) that forbids it from specifying what to change, where, why, or how; it
may only ask for further refactoring. Any difference between $V_{t-1}$ and $V_t$
is therefore the coding agent's own decision to keep modifying the code.
Each codebase lives on its own isolated branch and worktree, so the agent sees
only the target project, and the $T$ turns are committed in sequence to yield
$V_0, V_1, \ldots, V_{10}$.

\subsection{Datasets}
\label{subsec:datasets}

We use two families of minimal code that leave little room for meaningful refactoring, so additional edits largely amount to over-editing.

\paragraph{Algorithmic problems} The first benchmark family consists of 50 data structures and algorithms, randomly drawn from ~\cite{cormen2022introduction}, along with their implementations in the GeeksforGeeks reference solutions~\cite{geeksforgeeks}. The subjects cover common algorithmic topics, including searching and sorting, graph traversal and shortest paths, dynamic programming, balanced trees, hashing, string matching, and number theory. Each subject is a single self-contained Java file, intentionally small and already clean, ranging from 13 to 229 lines.

\paragraph{Object-oriented academic code} The second family is
object-oriented Java student coursework drawn from the Menagerie corpus of
graded assignment submissions~\cite{menagerie}. From this corpus, we select the
10 highest-graded submissions from the most recent years; each is a complete,
multi-class, multi-method program (roughly 16--34 source files) that complements
the algorithmic set with realistic object-oriented structure. Two authors
manually inspected the selected submissions to ensure their diversity and
quality.

\subsection{Agents and Models}
\label{subsec:models}

We study the two coding agents in widest use, Codex and Claude
Code, driving them through their command-line interfaces so that each turn is a
real agent session rather than a bare model call. Across the two agents we cover
five models: GPT-5.4, GPT-5.4-mini, and GPT-5.5 through Codex, and Claude
Opus~4.8 and Claude Sonnet~4.6 through Claude Code. We set the reasoning effort
to high for all five models.

\subsection{Software Quality Metrics}

We measure codebase quality at the initial state and after every iteration of each run. We compute these metrics using two static-analysis engines. SonarQube Community Build~25.1 (SonarJava~8.8, run on Java~17 via SonarScanner~8.1) provides cyclomatic and cognitive complexity, code smells, duplication density, the SQALE maintainability rating, and the technical-debt ratio. CK~0.7.0 provides class-level coupling between objects (CBO) and per-method cyclomatic complexity.

\begin{figure*}
    \centering
    \includegraphics[width=\linewidth]{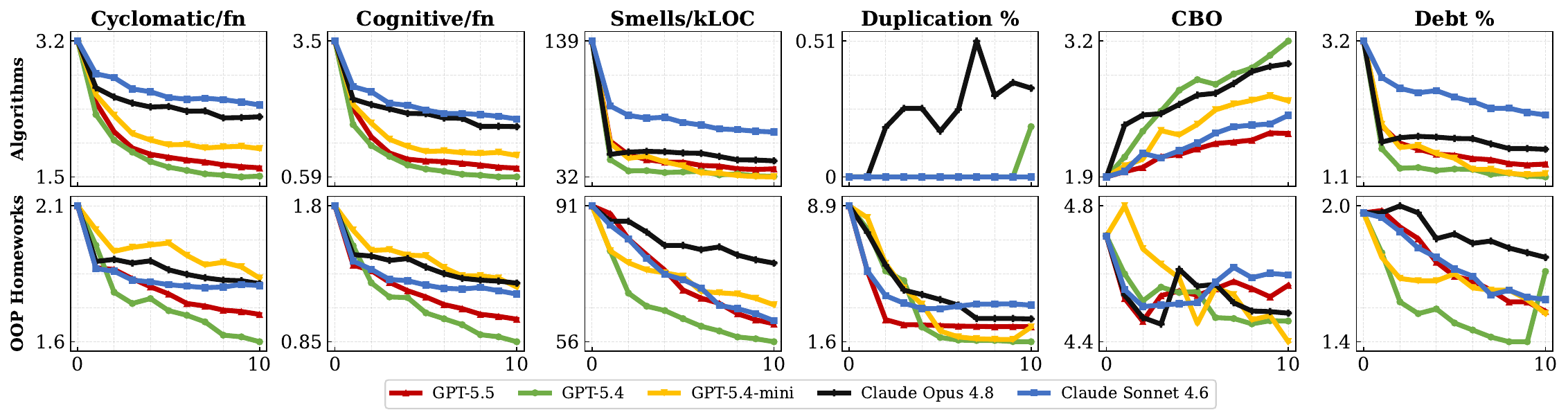}
    \caption{Quality metrics vs. editing turn, per branch and model. Cognitive complexity, code smells, and technical debt drop sharply after the first turn and then plateau, while coupling (CBO) continues to rise as editing continues, indicating that most improvement occurs in one pass, and later turns add code without quality gains.}
    \label{fig:grid-all-metrics}
\end{figure*}

Most quality improvement occurs in the first turn. Across branches, cognitive complexity, code smells, and technical-debt ratio all drop sharply after the first pass, while later turns provide only marginal additional gains (Fig.~\ref{fig:grid-all-metrics}). At the branch level, the first turn accounts for most of the net baseline-to-final improvement, and all models reach SonarQube's highest maintainability rating. Thus, the code is already near the best quality level the agent will achieve after one pass. Editing nevertheless continues after these quality metrics plateau. Later turns add code and increase coupling without corresponding quality gains. 

\subsection{Stopping Mechanism}

\label{subsubsec:stopping-taxonomy}
We treat a turn as a \emph{stop} when the agent finishes successfully and writes zero net lines:
$\Delta l_t = 0$ with an empty diff.
For each stop turn, we extracted the agent's reasoning trace and its reply to the user.
Two authors went through two rounds of collaborative qualitative coding~\cite{richards2018practical} and observed that three main categories emerged:
\emph{Refusal} is an explicit decline to act.
\emph{Optimum-awareness} claims the code is already optimum.
\emph{Questioning} redirects the request to the user and asks for a concrete goal before proceeding.
These labels are not mutually exclusive: a single turn may refuse, cite code quality, and ask for a goal.
Table~\ref{tab:stopping_taxonomy} summarizes the three categories, giving a representative matched fragment and its frequency for each. We report the full per-category prevalence in Section~\ref{sec:results}.

\begin{table*}[t]
\centering
\renewcommand{\arraystretch}{1.35}
\caption{AI Agent Stopping Communication taxonomy}
\label{tab:stopping_taxonomy}
\resizebox{\textwidth}{!}{%
\begin{tabular}{@{}p{0.15\textwidth}|p{0.43\textwidth} |p{0.29\textwidth}|p{0.10\textwidth}@{}}
\textbf{Category} & \textbf{Definition} & \textbf{Example} & \textbf{Count} (\%) \\
\hline
Refusal &
An explicit decline to act. &
\emph{``... I'm not going to refactor this again ...''} &
378 (83.3\%) \\
Optimum-awareness &
Claims the code is already as good as it should be and that further edits would be churn. &
\emph{``... The code is already well-refactored ...''} &
214 (47.1\%) \\
Questioning &
Redirects the request to the user and asks for a concrete goal before proceeding. &
\emph{``... If you have a concrete goal, share it.''} &
275 (60.6\%) \\
\hline

\end{tabular}%
}
\end{table*}

To apply the taxonomy at scale, we built a keyword detector using case-insensitive regular expressions for each category. A turn receives a label when any pattern for that category matches the agent reasoning or output message. We seeded the patterns from observed stop turns and tuned them on the \dsalg suite against 0 line changes turn. The initial detector achieved high recall ($0.993$) but low precision ($0.340$), producing 836 false positives for having generic phrases. Removing these patterns improved precision with little recall loss; on all 3{,}000 turns, the final detector flags 430 of the 454 stop turns, achieves precision $0.720$, recall $0.947$, $F_1 = 0.818$, and accuracy $0.936$. We favor high recall because the detector feeds $S_3$: if a stop claim is missed, the corresponding reasoning--action mismatch cannot be detected.

\subsection{Ablation Studies}
\label{subsec:ablations}
To test whether our findings depend on specific design choices, we conduct three ablations, each varying one factor while holding the others fixed: reasoning effort, prompt style, and codebase scale. We summarize the outcomes here and give full setups and results in the supplementary material~\cite{supp}.

\noindent\emph{Reasoning Effort.} This ablation tests whether over-compliance depends on the agent's deliberation budget. Holding the simulated developer, task protocol, interaction length, and measured signals fixed, we vary only the agent's effort setting, running GPT-5.5 on the Java binary-search subject with ten replicas at each level (\texttt{none}, \texttt{low}, \texttt{medium}, \texttt{high}, \texttt{xhigh}). \textit{The same pattern appears across the sweep: }the agent produces substantial edits instead of stopping, with higher effort producing more feature rollback and reimplementation. Over-compliance is thus not simply a consequence of insufficient reasoning.

\noindent\emph{Simulated Developer System Prompt Variation.} This ablation tests whether over-compliance depends on how the developer formulates the request. Keeping the protocol and target codebase fixed, we vary the prompting regime across three settings: a minimal static prompt, \texttt{refactor}; the novice developer used in the main experiment; and a simulated expert developer issuing concrete refactoring directions, using real-world prompt patterns from developers with more than five years of experience. Under the static prompt, agents are often uncertain about the user's intent, with some recognizing the interaction as a coding test environment, which motivates the LLM-based simulated developer. \textit{Under the expert-style prompt, pre-stop churn increases rather than decreases, and rollback becomes more pronounced.} Over-compliance is thus not an artifact of one novice formulation; it persists when the request appears more realistic and actionable.

\noindent\emph{Codebase Scale.} This ablation tests whether the behavior is specific to small programs. Keeping the protocol and 10-turn horizon fixed, we replace the benchmark subjects with four large open-source Java repositories from the GitHub top-100 by stars: DBeaver, Elasticsearch, Guava, Spring~Boot, and Termux, spanning tens to hundreds of thousands of lines across thousands of files. \textit{The core pattern persists: agents continue editing across turns, edits remain large, and fragile stopping reappears.} Although the normalized edit proportion is smaller relative to repository size, single turns still introduce changes on the order of thousands of lines. Over-compliance is thus not limited to small benchmarks; it also appears in production-scale repositories where refactoring is more open-ended and a clear completion point is harder to establish.

Across all three ablations, the conclusion is consistent: increasing reasoning effort, changing the simulated user, and scaling to larger codebases do not remove the tendency to continue editing code that warrants no further change.

\section{Results}
\label{sec:results}

\subsection{How Many Turns for Agents to Stop Refactoring?}
\begin{figure}
    \centering
    \includegraphics[width=\linewidth]{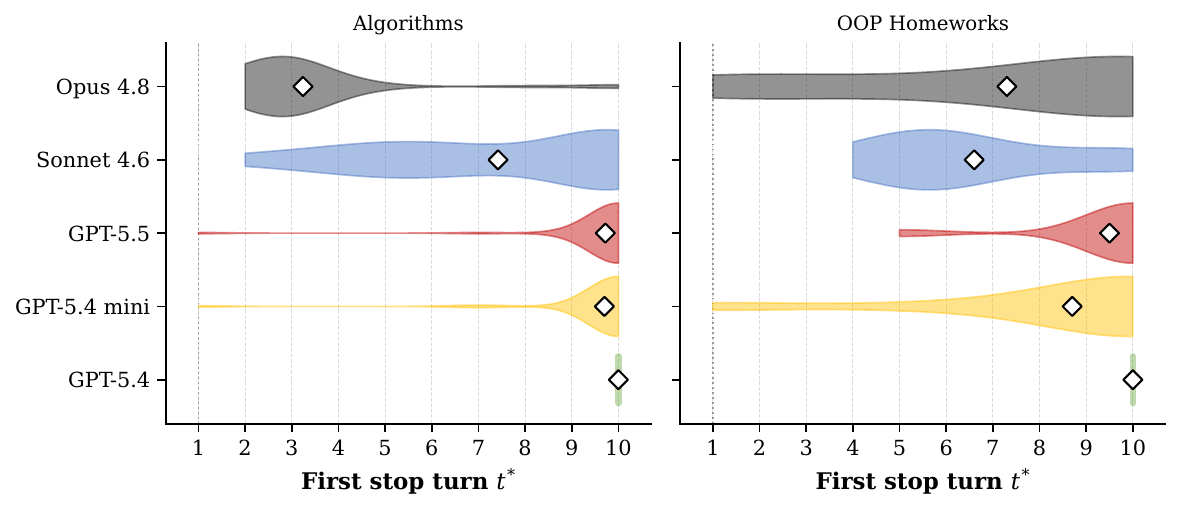}
    \caption{Distribution of the first-stop turn $t^{*}$. For each run,
$t^{*}$ is the earliest turn with no edit (runs that never stop are set to
$t^{*}=10$). Violins show the per-run spread by model, and white diamonds mark
per-model means, for the \dsalg (left) and \dsoop (right) suites. An agent that
respects an already-optimized codebase should stop at the first turn.}
    \label{fig:ts-stop_turn}
\end{figure}

Our setup repeatedly asks each agent to refactor a codebase, for up to ten turns. The most basic question is whether agents ever stop editing, and if so, how soon. Across the full experiment, only 454 of 3{,}000 turns result in no code edit, spanning 60 codebases, five models, and ten turns per run. We therefore examine $t^*$, the first turn at which an agent stops editing the codebase.

Figure~\ref{fig:ts-stop_turn} shows the distribution of $t^*$ across all models and both codebase suites. A clear behavioral divide emerges between the Claude and GPT model families. On the 50 algorithm codebases, Claude Opus 4.8 reaches the first stopping point earliest among all models, with a mean $t^*$ of approximately 3, followed by Claude Sonnet 4.6 with a mean near 5. In contrast, the GPT models tend to stop substantially later, clustering near turn 10, the final turn in our experiment. A similar pattern holds on the Java homework codebases. Claude Opus 4.8 stops later than on the algorithm codebases, with a mean $t^*$ around 7, while Claude Sonnet 4.6 stops at a similar point, around 6.6. By contrast, all three GPT models again concentrate near the end of the interaction horizon.

These results expose a systematic failure of editorial restraint across all models. The expected behavior is that the agent should either decline to make changes on the first turn, yielding $t^* = 1$, or apply at most one genuinely warranted cleanup pass and then stop, yielding $t^* = 2$. Instead, all models persist well beyond this threshold, repeatedly modifying the code in response to continued user requests.

Rather than pushing back on a request that no longer adds clear value, agents continue to accommodate it, producing edit churn that is overwhelmingly incidental rather than a meaningful improvement.

\subsection{How Many Lines Do They Change Before Stopping?}
\begin{figure*}
    \centering
    \includegraphics[width=\linewidth]{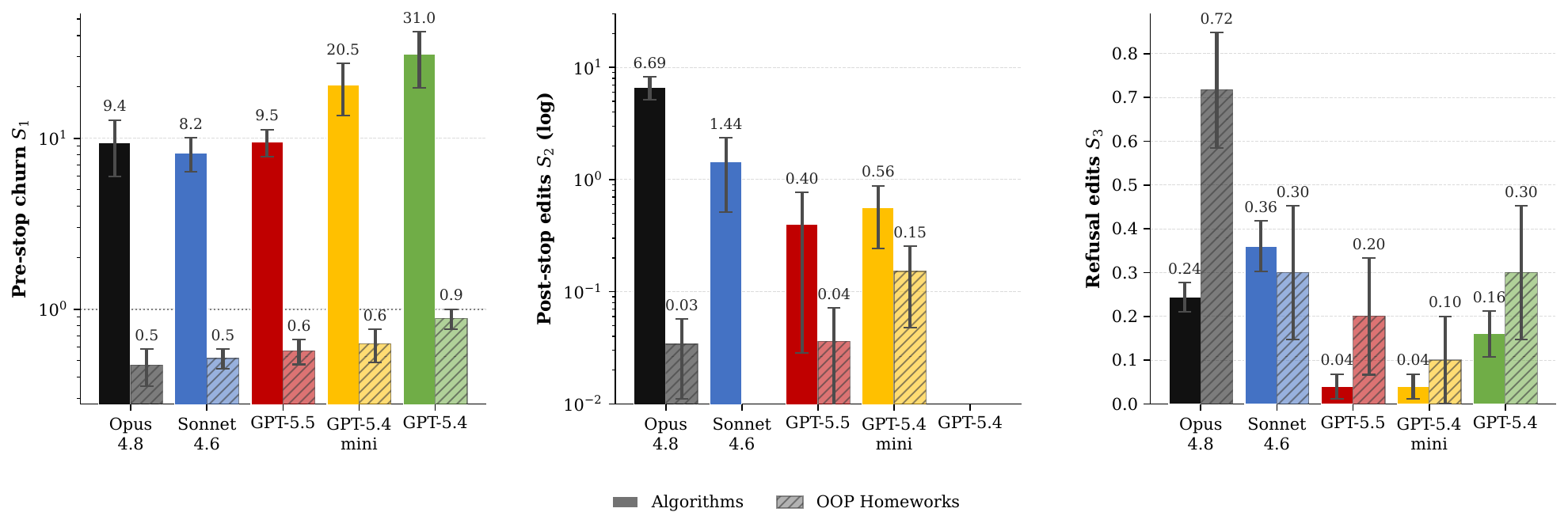}
\caption{Editing volume and over-compliance, by model on the \dsalg and \dsoop
suites (log scale; bars show mean $\pm$ SEM, dots are individual runs). $S_1$:
pre-stop churn, the lines changed before the first stop, normalized by the
original codebase size $L_0$; values above the $1\times$ dotted line mean the
agent rewrote more than the entire codebase before stopping. $S_2$: post-stop
modification, the lines changed across all turns after the first stop, also
normalized by $L_0$. $S_3$: the fraction of turns in which an agent edits the
code in the same turn it states no further change is needed.}
    \label{fig:s2s3-poststop-churn}
    \label{fig:s1-churn}
\end{figure*}

Figure~\ref{fig:s1-churn} presents the pre-stop churn \(S_1\), defined as the total number of lines changed before the agent's first stop, which is normalized by the original codebase size $L_0$. Values are averaged across the 50 codebases and shown on a log scale.

The volume of unnecessary editing is substantial across all models, but varies sharply by model family.
On the Algorithms benchmark, the lowest churn is observed for Claude Sonnet 4.6, at $8.22\times$ the original codebase size, followed closely by Claude Opus 4.8 at $9.41\times$ and GPT-5.5 at $9.52\times$.
Churn increases substantially for the remaining GPT models: GPT-5.4-mini reaches $20.52\times$, while GPT-5.4 peaks at $30.97\times$.
On the \dsoop benchmark, the same ordering largely holds, but at a smaller normalized scale, ranging from $0.47\times$ for Claude Opus 4.8 to $0.88\times$ for GPT-5.4. This lower magnitude is partly explained by the larger initial size of the Java homework codebases, whose original snapshots contain between 1,395 and 2,065 lines of code.
Whether measured as several times the codebase or a large fraction of it, the volume of change is far out of proportion to any plausible improvement to code that was already clean.

\subsection{How do they maintain stopping behavior?}
When an agent does decline, we examined what it actually says. Across the 454 turns in the \dsalg and \dsoop suites where an agent changed zero lines, the decline language falls into three recurring categories. \emph{Refusal} is an explicit decline to act, present in 378 of these turns, as in ``I won't make changes just because the request is repeated.'' \emph{Optimum-awareness} asserts that the code is already as good as it should be and that further edits would be churn, present in 214 turns. \emph{Questioning} redirects the request back to the user and asks for a concrete goal before proceeding, present in 275 turns. These categories are not mutually exclusive: a single turn often refuses, justifies the refusal by appeal to the code's quality, and asks for a concrete objective all at once, and only 24 of the 454 turns showed none of the three signals.

A model that states no further changes are needed but still edits the code fails to align its actions with its own assessment. We examined this using two complementary signals: $S_3$, which captures same-turn inconsistency between a stop claim and actual editing behavior, and $S_2$, which captures the cumulative editing that occurs after the first stop.

The $S_3$ results show that this action mismatch occurs frequently, especially for Claude models.
On the Algorithms benchmark, Claude Sonnet 4.6 edits despite a stop claim in $36.0\%$ of such turns, while Claude Opus 4.8 does so in $24.3\%$ of such turns.
These rates are higher than those of GPT-5.5 and GPT-5.4-mini, both at $4.0\%$, and GPT-5.4 at $16.0\%$.
The gap becomes larger on the Java homework benchmark, where Claude Opus 4.8 edits despite a stop claim in $71.7\%$ of such turns, more than twice the rate of the next-highest model.
This behavior represents a particularly problematic form of over-compliance: the agent signals awareness that no further edits are warranted, yet still modifies the code in response to the repeated user request.

Complementing this result, $S_2$ measures the total number of line changes accumulated after the first stop across all subsequent turns.
This metric captures not only whether a model later relapses, but also the magnitude of the resulting post-stop churn.
On the Algorithms benchmark, Claude Opus 4.8 accumulates the largest post-stop footprint of any model, at $6.69\times$ the original codebase size.
This is more than four times Claude Sonnet 4.6 ($1.44\times$) and nearly seventeen times GPT-5.5 ($0.40\times$).
GPT-5.4 shows no measurable post-stop churn because it never stops before the 10-turn horizon across the Algorithms experiments.
This contrast is especially important: Claude Opus 4.8 has the earliest mean first-stop turn on Algorithms, $t^* = 3.24$, making it the closest model to the desired early stopping behavior.
Yet it also produces the most editing after that stop.
When refactoring requests continue beyond $t^*$, Opus fails to maintain its initial restraint and resumes editing more aggressively than any other model.

Taken together, \(S_3\) and \(S_2\) show that early stopping is necessary but not sufficient: agents must also translate that recognition into action and remain robust to continued user pressure.

\subsection{What Are the Characteristics of These Changes?}
To characterize the substance of these changes, we apply RefDiff~\cite{silva2020refdiff} to classify modifications as matched structural refactorings, unmatched node additions, or unmatched node deletions.
We then examine whether edits exhibit cyclical add/remove behavior and whether they concentrate in the same regions across turns.
\begin{figure*}
    \centering
    \begin{subfigure}[t]{0.49\linewidth}
        \centering
        \includegraphics[width=\linewidth]{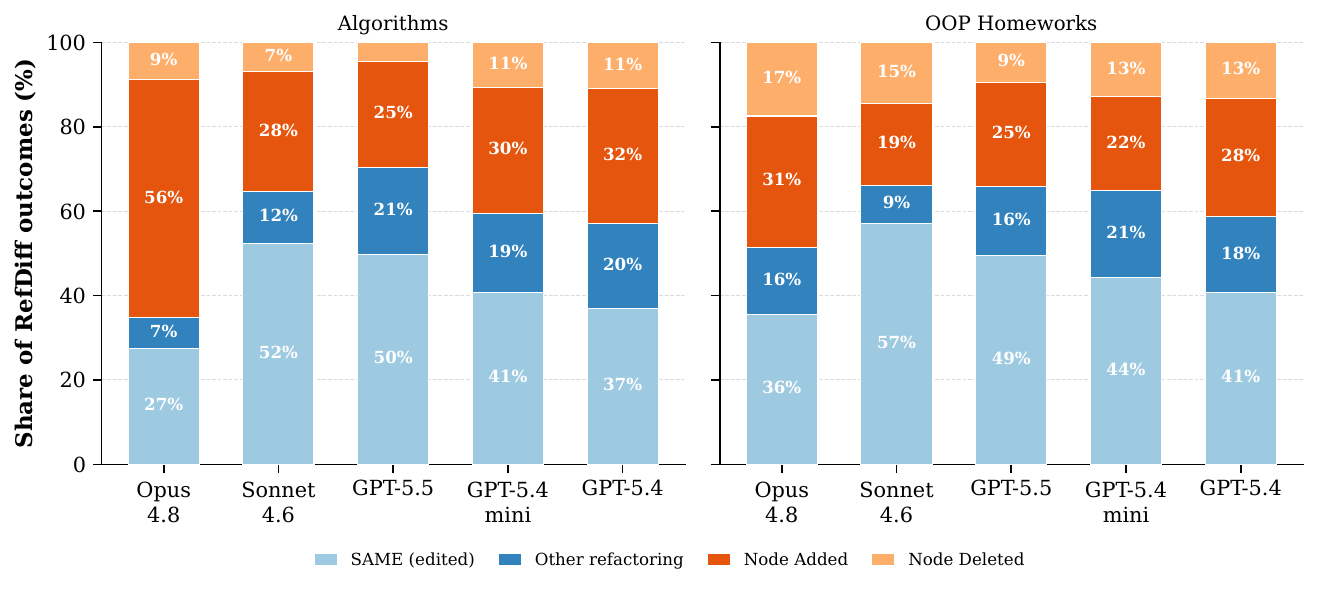}
        \caption{}
        \label{fig:refdiff-composition}
    \end{subfigure}
    \hfill
    \begin{subfigure}[t]{0.49\linewidth}
        \centering
        \includegraphics[width=\linewidth]{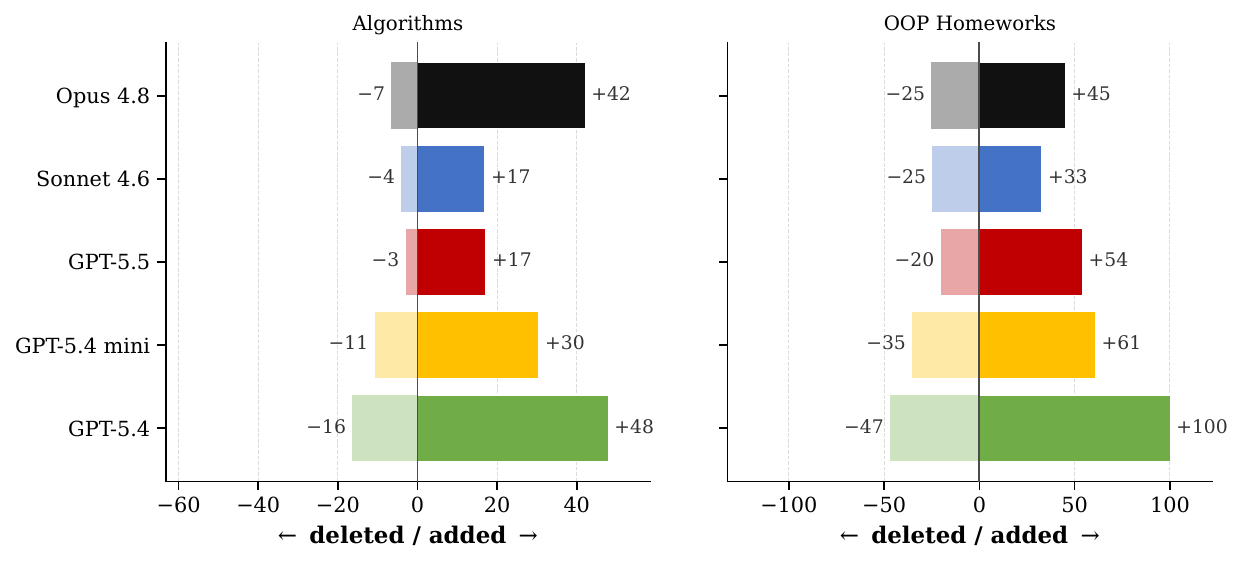}
        \caption{}
        \label{fig:inflation}
    \end{subfigure}
    \caption{a) RefDiff outcome composition per model (100\%-stacked), for the
        \dsalg and \dsoop suites. \texttt{SAME (edited)} marks code matched to an
        existing structural node but not identified as a catalogued refactoring;
        \texttt{Other refactoring} groups all named operations such as rename,
        move, extract, and inline; the remaining share is unmatched node
        additions and deletions. b) Node-level code inflation. For each model, mean nodes added
        (right) versus nodes deleted (left) per codebase, on the \dsalg and
        \dsoop suites. Every model adds far more structure than it removes.}
    \label{fig:structural-overview}
\end{figure*}

\begin{table}[t]
\centering
\renewcommand{\arraystretch}{1}
\caption{Mean \texttt{Extract} (growth) versus \texttt{Inline} (consolidation)
operations per codebase, by model, and suite.
Every model expands far more than it consolidates.}
\label{tab:extract-vs-inline}
\resizebox{\linewidth}{!}{%
\begin{tabular}{@{}ll rrr@{}}
\toprule
\textbf{Model} & \textbf{Suite} & \textbf{Extract} & \textbf{Inline} & \textbf{Ratio} \\
\midrule
Claude Opus 4.8   & \dsalg & 1.4  & 0.1 & 14.0$\times$ \\
                  & \dsoop & 7.2  & 0.2 & 36.0$\times$ \\
\addlinespace
Claude Sonnet 4.6 & \dsalg & 2.7  & 0.5 & 5.4$\times$  \\
                  & \dsoop & 3.5  & 1.5 & 2.3$\times$  \\
\addlinespace
GPT-5.5           & \dsalg & 8.0  & 0.3 & 26.7$\times$ \\
                  & \dsoop & 22.6 & 0.1 & 226.0$\times$ \\
\addlinespace
GPT-5.4-mini      & \dsalg & 6.3  & 1.2 & 5.3$\times$  \\
                  & \dsoop & 14.5 & 0.7 & 20.7$\times$ \\
\addlinespace
GPT-5.4           & \dsalg & 11.4 & 1.1 & 10.4$\times$ \\
                  & \dsoop & 27.5 & 0.5 & 55.0$\times$ \\
\bottomrule
\end{tabular}
}
\end{table}

\paragraph{Most edits are not structural refactoring}
Figure~\ref{fig:refdiff-composition} breaks down RefDiff outcomes by category.
Across all models, the dominant category is \texttt{SAME (edited)}, which denotes code that RefDiff matches to a prior structural node but does not identify as a deliberate refactoring operation such as rename, move, extract, or inline.
On the Algorithms benchmark, this category accounts for $27.5\%$ of outcomes for Claude Opus 4.8 and up to $52.4\%$ for Claude Sonnet 4.6. By contrast, genuine structural operations, grouped under \texttt{Other refactoring}, account for only $7.2\%$ to $20.6\%$ of outcomes across models.
The remaining outcomes are largely unmatched node additions, which we analyze next.
These results indicate that repeated refactoring prompts mostly produce ordinary code edits rather than disciplined structural transformations.

\paragraph{Repeated refactoring requests produce code inflation}
A second pattern compounds this concern: models systematically add more code than they remove.
Figure~\ref{fig:inflation} shows the divergence between added and deleted nodes for each model.
On the Algorithms benchmark, Claude Opus 4.8 adds $42.1$ nodes on average while deleting only $6.6$, a ratio of roughly $6.4{:}1$.
Claude Sonnet 4.6 adds $16.9$ nodes while deleting $4.1$, a ratio of $4.1{:}1$.
The imbalance is largest for GPT-5.4, which adds $47.9$ nodes while deleting $16.4$.
The same pattern holds on the \dsoop benchmark, where every model again adds substantially more code than it removes.
This code inflation increases the surface area of the codebase and may create downstream review and maintenance burden. 

\paragraph{Growth without consolidation.}
The same asymmetry appears at the level of specific refactoring operations.
Table~\ref{tab:extract-vs-inline} compares the mean number of \texttt{Extract} operations, which introduce new functions or classes, against \texttt{Inline} operations, which consolidate code by folding extracted elements back into their callers.
On the Algorithms benchmark, every model performs substantially more Extracts than Inlines.
GPT-5.5 performs roughly $26\times$ more Extracts than Inlines ($7.9$ vs.\ $0.3$, combining \texttt{EXTRACT} and \texttt{EXTRACT\_MOVE}), Claude Opus 4.8 performs roughly $13\times$ more ($1.3$ vs.\ $0.1$), and Claude Sonnet 4.6 performs roughly $4.8\times$ more ($2.4$ vs.\ $0.5$).
The pattern is even stronger on the \dsoop benchmark, where GPT-5.5 performs over $200\times$ more Extracts than Inlines, and GPT-5.4 performs over $50\times$ more.
No model approaches a balanced ratio between expansion-oriented and consolidation-oriented operations in either benchmark.

At the operation level, this confirms the node-count pattern: repeated refactoring requests produce predominantly growth-oriented edits rather than balanced expansion and simplification.
This reinforces the maintainability cost of over-compliant behavior:
agents respond to continued user pressure by creating more code and more structure, even when there is no clear evidence that the existing design requires it.

\begin{figure*}
    \centering
    \includegraphics[width=\linewidth]{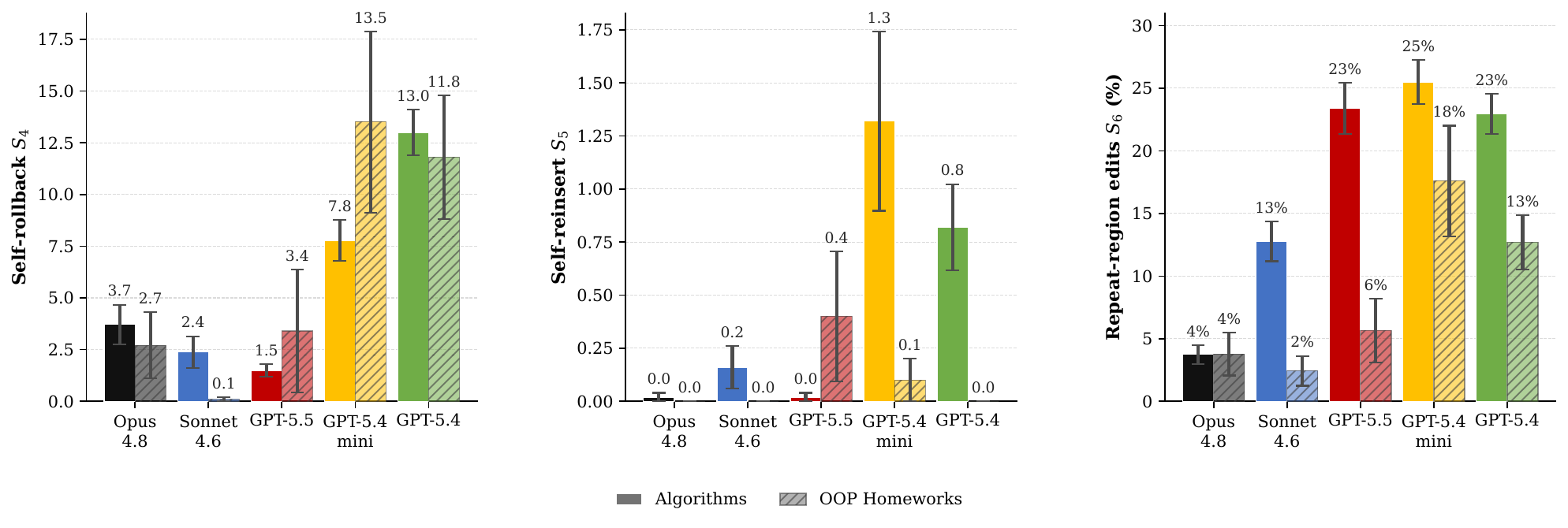}
    \caption{Structural edit behavior across models on the \dsalg and \dsoop suites. Edit instability: $S_4$ counts elements an agent removes after adding them on an earlier turn, and $S_5$ counts elements it re-adds after deleting them (bars show per-model means). Edit locality ($S_6$): the share of edits landing in a region the model modified on an earlier turn (mean $\pm$ SEM).}
    \label{fig:s4s5instability}
    \label{fig:s6locality}
\end{figure*}
\paragraph{Agents exhibit self-cancelling instability}
Closer inspection of edit sequences shows that models do not merely add code; they also sometimes contradict their own earlier edits.
We use $S_4$ to capture cases where a model removes an element that was previously added, and $S_5$ to capture the reverse pattern, where a model reintroduces an element that it had previously deleted.
On the Algorithms benchmark, $S_4$ is lowest for GPT-5.5 ($1.5$) and Claude Sonnet 4.6 ($2.38$), followed by Claude Opus 4.8 ($3.72$) as shown in Figure~\ref{fig:s4s5instability}.
It rises sharply for GPT-5.4-mini ($7.78$) and peaks for GPT-5.4 ($13.0$), nearly $9\times$ the GPT-5.5 value.
The reverse pattern, $S_5$, is rarer across all models but follows a similar trend.
GPT-5.4-mini reintroduces previously deleted elements at a rate of $1.32$, the highest among all models, compared with $0.02$ for both Claude Opus 4.8 and Claude Sonnet 4.6.

These self-cancelling add/remove loops suggest that agents are not consistently converging toward a stable target state.
Instead, repeated requests can induce oscillatory behavior, in which agents undo and redo their own prior work across turns.
This represents an instability dimension of over-compliance:
rather than maintaining a coherent judgment about what the code requires, the agent's edits continue to shift as the user repeats the request.

\paragraph{Edit locality differs across model families}
Finally, we examine the spatial distribution of edits using $S_6$, the percentage of edits that occur in a region the model has already modified in a previous turn.
Figure~\ref{fig:s6locality} shows a clear separation between model families.
On the Algorithms benchmark, Claude Opus 4.8 re-edits the same region in only $3.74\%$ of cases, and Claude Sonnet 4.6 does so in $12.78\%$ of cases.
By contrast, all three GPT models re-edit the same region between $22.9\%$ and $25.5\%$ of the time, roughly 2$\times$ Sonnet and more than 6$\times$Opus.
The same ordering holds on the \dsoop benchmark, where Claude models remain below $4\%$, while GPT models range from $12.7\%$ to $17.6\%$.

This difference persists even after accounting for the fact that GPT models generally take more turns before stopping, which naturally gives them more opportunities to revisit the same location.
After normalizing by $t^*$, Claude Opus 4.8 has the lowest $S_6/t^*$ ratio on both benchmarks ($1.15$ on Algorithms and $0.52$ on \dsoop), while GPT-5.4-mini has the highest ratio on both ($2.63$ and $2.02$, respectively).
Thus, GPT models do not revisit the same regions merely because they continue for more turns; they are also more likely, per turn, to return to previously modified locations.
In contrast, Claude models distribute edits across a wider range of locations as refactoring requests continue.

These are two distinct forms of over-compliance, both undesirable in already-optimized code: GPT models concentrate churn in the same regions, while Claude models, especially Opus, broaden the footprint of unnecessary change across the codebase.

\section{Discussion}
\label{sec:discussion}
Sycophancy is a documented behavior of large language models: they agree
with stated views, endorse weak arguments, and abandon correct answers
under pushback~\cite{sharma2024towards, perez2022sycophancy}. The
natural-language literature attributes this to preference optimization,
and to RLHF in particular, which rewards responses a rater approves of
rather than correct ones~\cite{ouyang2022instructgpt, shapira2026rlhf}.
We establish the consequences of this tendency in software development,
then discuss how they can be detected and mitigated.

\subsection{Consequences of Sycophancy in Software}
\label{sec:consequences}
A sycophantic model fails differently across domains. In a coding agent, the unit of harm is an \emph{edit},
and its cost is durable: a misleading sentence can be forgotten, but an
unnecessary edit must be reviewed and, once merged, maintained. Our
results surface four such consequences.

\emph{Unnecessary editing.}
Agents keep modifying code that is already sound. No model declined on
the first turns, and even the most restrained accumulated pre-stop churn
of roughly $8\times$ the original codebase, rising above $30\times$ for
the worst (Fig.~\ref{fig:s1-churn}). This churn is not disciplined
restructuring: across models, $27.5$--$52.4\%$ of edits are
\textsc{same}-edited regions RefDiff matches to an existing node but
cannot classify as any cataloged refactoring
(Fig.~\ref{fig:refdiff-composition}), so the dominant activity is
incidental rewriting. Such edits enlarge diffs, and review
effectiveness falls as diff size grows~\cite{bacchelli2013expectations};
each also carries a nonzero chance of introducing a defect into
previously clean code~\cite{nagappan2005use}.

\emph{Fragile stopping.}
Agents that recognize completion resume editing under continued
prompting. The contradiction is sharpest for the best-behaved model:
Claude Opus~4.8 reaches the earliest mean stop of any agent
($t^*\!=\!3.24$ on the algorithm suite), yet edits in $71.7\%$ of the
turns where it declares no change is needed on the OOP suite, and
accumulates the largest post-stop churn of any model
($S_2\!=\!6.69\times$; Fig.~\ref{fig:s2s3-poststop-churn}). A stopping
decision is thus not a stable assessment but a state the next request
can overturn. An agent whose ``no'' dissolves under repetition cannot be
trusted to defend a correct one.

\emph{Code inflation.}
Repeated requests drive growth, not simplification. Every model adds far
more structure than it removes---Claude Opus~4.8 adds $42$ nodes per
codebase while deleting $7$ (Fig.~\ref{fig:inflation})---and extracts far
more than it inlines, by up to $26\times$
(Table~\ref{tab:extract-vs-inline}). Individually, these look like
ordinary refactorings, but their cumulative effect raises structural
complexity without functional gain which reduces
maintainability~\cite{lehman2005programs, gibson1989system}.

\emph{Oscillation.}
Agents remove code they added a few turns earlier and reintroduce code
they deleted, with rolled-back additions reaching $13$ per codebase for
GPT-5.4 (Fig.~\ref{fig:s4s5instability}). Such self-canceling edits show
the agent is not converging on a target state but reacting to each
request in turn---over-compliance in its purest form.

\subsection{Detecting and Mitigating Over-Compliance}
\label{sec:mitigation}
These consequences are measurable, and thus actionable. We note three
complementary routes.

\subsubsection{Extending the behavioral signals}
Our signals are lightweight. Pre- and post-stop churn ($S_1$, $S_2$) are
computable from version-control history alone and fit into
pull-request or CI workflows~\cite{nagappan2005use}, while the
stop-then-edit discrepancy ($S_3$) needs only the transcript alongside
the diff. We intend $S_1$--$S_6$ as a starting framework rather than a
closed taxonomy.

\subsubsection{Adapting natural-language mitigations}
Because the behavior is model-level, conversational-sycophancy
mitigations transfer in principle. Synthetic-data interventions that
reduce capitulation under pushback~\cite{wei2023simple} could be
retargeted so that ``keep refactoring'' on clean code is met with a held
position. Our signals also supply a training target the natural-language
setting lacks: excess churn can be penalized and appropriate stopping
rewarded during alignment, and $S_3$ can align what an agent does with
what it says. For coding agents, rewarding \emph{when to stop} may matter
as much as rewarding successful generation.

\subsubsection{Building detectors for software-engineering}
The strongest opportunity is unique to our setting: unlike a
natural-language reply, a code edit has formal semantics, so we can ask
directly whether an edit changed anything that matters. Three families
of analysis address three distinct questions.

\emph{Did the edit change the behavior?}
On already-correct code, the honest answer after convergence is ``no,''
so a behavior-preserving edit under a repeated vague request is a
candidate for over-compliance. Even bounded or test-based equivalence
suffices to flag a turn as behaviorally inert. Differential
testing~\cite{mckeeman1998differential} is the practical, decidable
form---run $V_{t-1}$ and $V_t$ on shared inputs and report a turn whose
outputs never diverge---and it scales where symbolic methods do not.
Dynamically mined invariants offer a complementary check: tools such as
Daikon~\cite{ernst2001dynamically} infer properties that should be
preserved across an edit, so one that breaks an invariant on previously
correct code has introduced a regression.

\emph{Did the edit improve any quality property?}
Some behavior-preserving edits improve readability, so preservation
alone cannot flag over-compliance. The second test is whether the edit
improves a measurable quality property. Established complexity and
coupling measures---cyclomatic complexity, coupling, and
cohesion~\cite{mccabe1976complexity, chidamber1994metrics}---can gate an
edit: a turn that alters code while leaving these metrics flat or worse
is inflation, not improvement. This catches the extract-without-inline
pattern we observed: unless an extraction reduces complexity or
coupling, it does not improve the code.

\emph{How far does the edit reach?}
The third question is locality and impact, which dependence-based
analyses answer. Program slicing~\cite{weiser1984slicing} bounds the
computations an edit can affect; edits whose backward slices never reach
an observable output are inert, however many lines they touch. Slice
overlap across turns gives a dependence-graph analog of our
region-recurrence signal $S_6$, replacing textual co-location with
semantic dependence. Because all three families key on semantics rather
than surface text, they are robust to the cosmetic
rewrites---renamings, reorderings---that dominate our non-structural
edits.

Together, these techniques let developers and tool builders repurpose
equivalence checking, invariant detection, metric gating, and slicing as
over-compliance detectors in code review and CI, catching unnecessary
edits exactly where they cost developer time.
\section{Related Work}
\label{sec:related}

\subsection{Failure Modes of Language Models}

Large language models exhibit well-documented failure modes beyond fluent generation. Hallucination is the most widely studied: models can produce plausible but unsupported content, with causes spanning data, modeling, decoding, and alignment choices~\cite{huang2025hallucination, alansari2026hallucination}. Closest to our work is sycophancy. Sharma et al. show that assistants trained with human feedback often agree with a user's stated view, endorse weak arguments, and revise correct answers after user pushback; they also show that human preference data can reward this behavior~\cite{sharma2023sycophancy}. Related work finds similar tendencies across free-form tasks, suggesting that sycophancy is not limited to a single prompt format~\cite{perez2022sycophancy, wei2023synthetic}. This literature studies sycophancy primarily in natural-language outputs. We study its analogue in coding agents, where the model's response is not only text but also a modification to a software artifact.

\subsection{Failure Modes of Coding Agents}

A separate line of work studies how coding agents fail in software tasks. One cluster focuses on underspecified instructions. Vijayvargiya et al. introduce Ambig-SWE, an ambiguous variant of SWE-bench, and show that agents often fail to recognize when a request lacks enough information; asking clarifying questions recovers much of the lost performance~\cite{vijayvargiya2025ambigswe}. A second cluster studies instruction conflict and adversarial compliance. The Compliance Paradox shows that graders tuned for helpfulness may follow hidden directives inserted into submissions rather than judge the code correctly~\cite{sahoo2026compliance}, while robustness studies show that small prompt perturbations can steer models toward flawed outputs~\cite{liu2025adversarial}. Other work identifies specification drift and prompt-biased edits as recurring agent failure modes~\cite{nanda2026wink,tambon2025bugs}.

These studies usually treat failure as an incorrect, unsafe, or misaligned output within a single exchange. Over-compliance is different: each individual edit may appear plausible, but the failure emerges as the agent continues to modify code when no change is warranted. Existing coding-agent benchmarks primarily reward successful issue resolution and do not reward the absence of unnecessary change~\cite{jimenez2023swebench, ferrag2026agents}. As a result, they do not directly measure whether an agent can stop editing when stopping is the correct behavior.

\subsection{Refactoring as an Open-Ended Task}

Refactoring is behavior-preserving change without a natural stopping point, making it a useful setting for studying over-compliance. Recent benchmarks show that agents depend heavily on task specification to determine what refactoring is appropriate.
CODETASTE finds that models perform better when told exactly what to change, but struggle to infer the intended refactoring from vague improvement requests~\cite{thillen2026codetaste}. Classical refactoring work also treats unnecessary change as a cost: Ouni et al. formulate refactoring recommendation as a multi-objective search problem that minimizes the number of changes while preserving design semantics~\cite{ouni2016refactoring}. This makes refactoring a useful test case for whether agents recognize when no edit is warranted, rather than continuing to generate changes simply because the user repeats the request.


This literature shows that extra edits are not benign. We extend this line of work by identifying a concrete source of unnecessary change: agents that continue to comply with repeated refactoring requests even when the code no longer warrants modification.

\section{Limitations}

Two technical choices constrain our signals. We rely on RefDiff~2.0 to recover
structural changes at the granularity of named code elements (classes,
interfaces, methods), so statement-level and local-variable edits are invisible
to signals S4--S6. This coarseness is deliberate: method-level granularity is the
natural unit of refactoring, is consistent across models, and avoids conflating
refactoring with routine maintenance. Separately, LLM outputs are stochastic, so
a single run per codebase introduces variance; we control for this across 50
algorithm codebases and 10 OOP subjects, where consistent patterns across all 60
make it unlikely that run noise drives the findings.

Our design also limits generalizability. The core dataset of 60 Java codebases
(50 algorithmic problems, 10 object-oriented assignments) is modest by
corpus-mining standards, but five models over 10 turns each yield 3{,}000
interaction turns under a controlled setup, following the trajectory-based
evaluation paradigm of prior work~\cite{yang2024swe,jimenez2024swe}. Replicating
four full open-source projects reproduce the same over-compliance patterns at
attenuated magnitude, showing the effect is not an artifact of small or
pedagogical code. Subjects are Java-only because RefDiff's Java support is most
mature and both dataset families are naturally Java-based; whether
over-compliance manifests differently in dynamically typed or functional
languages is open. Our simulated user, a vibe-coder pushing for more refactoring
with no goal or stopping criterion, is a worst-case rather than a representative
scenario, yet neither the LLM-based prompt-modifier ablation nor the expert-prompt
ablation (using real phrasings from professional developers) suppresses the
behavior, supporting the claim that over-compliance originates in model priors
rather than prompt artifacts. Finally, we do not verify whether the agent-produced
code remains functionally correct after successive refactoring turns; that link
is an important direction for follow-on work.

\section{Conclusion}
AI coding agents are designed to help developers modify code, but little is known about whether they know when to push back on the user. We presented the first controlled study of over-compliance in coding agents, evaluating five state-of-the-art models across 60 codebases under repeated refactoring requests. All models continued editing code even when no further changes were needed, producing substantial unnecessary churn, and often reversed their own stopping decisions when users kept asking. Most edits were superficial rewrites that increased code size without improving quality. These results show that over-compliance is a systematic behavior that grows with continued interaction. We also introduce six lightweight signals that can help detect and reduce unnecessary editing in practice. As coding agents become more autonomous, knowing when not to modify code should be treated as a core capability.


\bibliographystyle{IEEEtran}
\bibliography{main}

\end{document}